# Whom to Ask? Jury Selection for Decision Making Tasks on Micro-blog Services

Caleb Chen Cao   Jieying She   Yongxin Tong   Lei Chen
Department of Computer Science and Engineering
The Hong Kong University of Science and Technology
Clear Water Bay, Kowloon, Hong Kong SAR, PR China
{caochen,jshe,yxtong,leichen}@cse.ust.hk

## ABSTRACT

It is universal to see people obtain knowledge on micro-blog services by asking others decision making questions. In this paper, we study the Jury Selection Problem(JSP) by utilizing crowdsourcing for decision making tasks on micro-blog services. Specifically, the problem is to enroll a subset of crowd under a limited budget, whose aggregated wisdom via Majority Voting scheme has the lowest probability of drawing a wrong answer(Jury Error Rate-JER).

Due to various individual error-rates of the crowd, the calculation of JER is non-trivial. Firstly, we explicitly state that JER is the probability when the number of wrong jurors is larger than half of the size of a jury. To avoid the exponentially increasing calculation of $JER$, we propose two efficient algorithms and an effective bounding technique. Furthermore, we study the Jury Selection Problem on two crowdsourcing models, one is for altruistic users($AltrM$) and the other is for incentive-requiring users($PayM$) who require extra payment when enrolled into a task. For the $AltrM$ model, we prove the monotonicity of JER on individual error rate and propose an efficient exact algorithm for JSP. For the $PayM$ model, we prove the NP-hardness of JSP on $PayM$ and propose an efficient greedy-based heuristic algorithm. Finally, we conduct a series of experiments to investigate the traits of JSP, and validate the efficiency and effectiveness of our proposed algorithms on both synthetic and real micro-blog data.

## 1. INTRODUCTION

Crowdsourcing, partially categorized as human computation or social computation, is an emerging computation paradigm. It provides fundamental infrastructure to enable online users to participate certain tasks as intellectual crowds. Amazingly, the wisdom of crowds outperforms computer programs at tasks involving creativity, human natural interpretation and subjective comparison, etc. In current stage, typical crowdsourcing applications entail specially designed platforms, like Amazon MTurk, to enroll crowd workers, control task flow and aggregate answers. On such platforms, crowd workers select tasks according to their own interests and reward requirement; on the other side, tasks requesters publish their tasks and wait for crowd workers accept and complete them in a random manner.

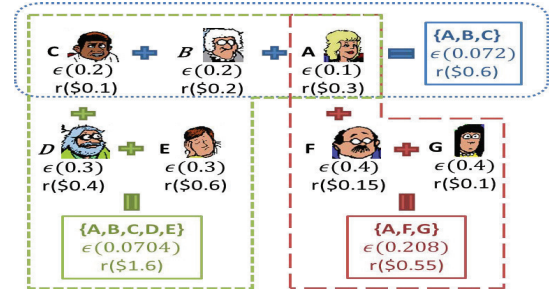

Figure 1: Is Turkey in Europe or in Asia?

Will the magic power of crowdsourcing be confined solely on specially designed platforms? Are crowd workers only appearing because of monetary rewards? The answer are both negative. In this paper, we will introduce a long-existing pattern of crowdsourcing on the platform of a micro-blog network, that is, gathering answers for decision making questions from micro-blog followers.

Micro-blog services are popular social media, featuring excellent brevity to broadcast observation of events and express users' opinions. This brevity is brought forward by the limited length of published content, e.g. 140 characters for Twitter, and a brief markup culture like "RT" and "@", which make it easy and even motivated for users to express their thoughts. The simplicity of use of micro-blog services encourages people to present their thoughts freely. Moreover, as mobile and web techniques advance, it becomes easier and easier for users to "tweet" via various ways. Besides its high accessibility, the huge population and diversity of users enable micro-blog services as a potential but powerful knowledge-base.

For the reasons above, micro-blog service is born a platform qualified not only for spreading message, but also for crowdsourcing particular tasks, e.g. answering decision making questions. Each day, people find it more and more convenient and reliable to seek answers from micro-blog users, for example, "Is it true that iPhone5 will come before August?" or "Is Doner Kebab available in Hong Kong?", etc. Such questions vary from minor problems as selection of dressing for a banquet to serious issues such as the prediction of macro markets trends. The magic point of such tasks on micro-blog networks is that question holders can actively choose their potential "workers" by simply mentioning them using the markup operator '@'. Later on "workers" return their product with the simple "Reply" button. Table 1 describes the genome of such tasks as crowdsourcing application.

**Table 1: Genome of Decision Making Tasks with Crowdsourcing [18]**

| Who | Why | What | How |
|---|---|---|---|
| Anyone | Altruism+Incentive | Decision | Group Decision |

Another type of formal application of such decision making tasks is the discerning of rumorous messages [4]. Rumors are spread on micro-blog network, such as political astroturf and spam advertising[24], etc. And it is very difficult to discover and identify them by automatic algorithms. The main reason is that most of rumors look seemingly the same as truth, or expressed with plenty of rhetoric or sarcasm. To discern such rumors is thus a typical decision making problem for online users, which has long been practiced by micro-blog users. In practice, micro-blog users are utilized to monitor and identify earthquake information during the disasters in Japan and Chile [27][4].

Following the same terminology in proposing judgement on court, which is one of the most typical decision making scenarios, in this paper we denote crowdsourcing "workers" as "jurors", and the "crowds" as "jury". Although "jurors" in a jury all wish to achieve the same goal(to answer the question correctly), each of them may make mistakes with a probability $\epsilon_i$. Meanwhile, some of them will not participate in such a task unless a certain incentive $r_i$ is offered. Not surprisingly, we hope to choose a most reliable and feasible subset of all possible "jurors" to vote on a question. Most reliable here means the possibility of giving wrong answer under majority rule is minimized. Formal definition of the problem will be given in Section 2, and a motivation example is given as follows.

**Motivation Example** Suppose we are given a decision task, with a set of candidate "jurors" $S$(i.e. all the users in Figure 1), we have to decide whom we should ask for an answer.

The first concern originates from the existence of defect, i.e. the probability of making mistakes among users in a jury. In Majority Voting(most classical Condorcet-like voting [9]), if more than half of the users vote wrongly, the jury gives wrong answer. For the example in Figure 1, if we choose C, D, and E, with error-rates 0.2, 0.3, and 0.3 respectively, as a crowd, the probability of getting a wrong answer from the entire crowd is

$$0.2 \cdot 0.3 \cdot 0.3 + (1-0.2) \cdot 0.3 \cdot 0.3 + 2 \cdot 0.2 \cdot (1-0.3) \cdot 0.3 = 0.174$$

This jury performs better than any individual of them does (e.g. 0.2 if only C is selected and 0.3 if select D or E is selected). Intuitively, we expect that the best jury comes from the best individuals. And indeed, with A, B, and C, the overall error-rate becomes 0.072, which is smaller than with C, D and E. What if the size of jury expands with two more individuals? After taking D, E into the jury, the error-rate becomes 0.0704, which is even lower. Following such intuition, when we take two more individuals F and G, we find that the error-rate climbs to 0.085, which is worse than that of the previously smaller jury with size 5. Regarding such cases, we are interested in the problem of selecting

**Table 2: Error-rate of Example in Figure 1**

| Crowd | Individual Error-rate | Jury Error-rate |
|---|---|---|
| C | 0.2 | 0.2 |
| A | 0.1 | 0.1 |
| C,D,E | 0.2,0.2,0.3 | 0.174 |
| A,B,C | 0.1,0.2,0.2 | 0.072 |
| A,B,C,D,E | 0.1,0.2,0.2,0.3,0.3 | 0.0703 |
| A,B,C,D,E,F,G | 0.1,0.2,0.2,0.3,0.3,0.4,0.4 | 0.0805 |
| A,B,C,F,G | 0.1,0.2,0.2,0.4,0.4 | 0.104 |

members for a jury with the lowest error-rate.

The second concern is about how to promote activity and productivity of the crowd. Financial rewards or other incentives(e.g. virtual commercial credits) are employed and proved to be effective. Incentive requirements vary among all the workers [13], so what if we cannot enroll the best jury due to a limited budget? In the example of Figure 1, because user D and E ask for rewards of $0.4 and $0.65 respectively, the sum of which already exceeds the $1 budget, the jury of A, B, C, D, and E cannot be formed. Should we give up D and E or should we take two cheaper but less reliable users F and G? The result in Table 1 shows that, in such settings, the smaller and cheaper jury with error-rate 0.072 will perform better than the larger but more expensive one with error-rate 0.104. This dilemma reveals the second concern: how to select the best jury with a limited budget.

In this paper, we propose a framework to form such crowdsourcing function on micro-blog services, and particularly investigate the power of crowds to tackle decision making tasks with higher quality. In general, we have such contributions:

1. We propose *AltrM* and *PayM* models to describe the behavior of crowd workers in crowdsourcing applications, and we formally propose the Jury Selection Problem(JSP) on both models;

2. We explicitly state the complexity of calculating Jury Error Rate and propose efficient algorithms with lower bounding criteria;

3. To solve JSP, we proved the monotonicity of JER within a fixed size of jury, and propose an efficient algorithm to tackle JSP on the *AltrM* model;

4. We prove that the Jury Selection Problem, under *PayM* Model, is NP-hard, and provide a polynomial heuristic algorithm to solve the problem.

5. We propose a method to retrieve users' individual error rates via constructing message forwarding graph and ranking the users.

## 2. MODELS AND PROBLEM DEFINITION

In this section, we first introduce the concept of jury and the fundamental voting scheme for a decision making task, as well as two crowdsourcing models for selecting a jury. Then we investigate the effect of careless jurors and the Jury Error Rate drawn from them. At the end of this section, we formally define the Jury Selection Problem.

### 2.1 Voting Scheme

Each online user of micro-blog services can serve as a juror, and the problem is how to select jurors and aggregate their distributed opinions, so that the wisdom of crowds is best utilized.

The term *jury* is used to denote a set of jurors that can make decisions on court, and here we borrow the concept for our decision making problem and redefine it formally as follows:

DEFINITION 1 (JURY). *A jury $J_n = \{j_1, j_2, \cdots, j_n\} \subseteq S$ is a set of jurors with size n that can form a voting.*

In most crowdsourcing applications, how to aggregate the wisdom of crowds is an important issue. The expected result may be a predicted value, a rank of several items, labels or annotations of to different items, etc. There are two main schemes for aggregation of wisdom of crowds. On one hand, synthesis, like market prediction, is in natural numerical and can be aggregated arithmetically on the individual quantities. On the other hand, for decision making tasks where there is "no natural way to 'average' the preference of individuals"[9], voting is a mainly adopted scheme. Voting is also considered as one of the most suitable aggregating mechanisms when intrinsic divergence exists among all individuals, but a hard group decision must be made after synthesis.

We thus give the definition of a *voting* as follows:

DEFINITION 2 (VOTING). *A voting $V_n$ is a valid instance of a jury $J_n$ with size n, which is a set of binary values.*

### 2.1.1 Majority Voting

A voting scheme defines how to aggregate a voting result so that a specific decision can be made. Specifically, we treat a voting scheme as a function defined on a *voting*(see Definition 2), and the output is a decision.

Aggregating the opinions on a decision making task resembles the procedure of drawing a consensus from jury in the court. One most natural and clearest mechanism to make a single decision is Majority Voting, which takes the opinion that is supported by more than half of the jurors. In order to give a clear answer, we assume the size of a jury for Majority Voting is **ODD**. Formally, we define Majority Voting as follows:

DEFINITION 3 (MAJORITY VOTING - MV). *Given a voting $V_n$ with size n, Majority Voting is defined as*

$$MV(V_n) = \begin{cases} 1 & if \sum j_i \geq \frac{n+1}{2} \\ 0 & if \sum j_i \leq \frac{n-1}{2} \end{cases}$$

### 2.1.2 Error-rate

The niche of collecting wisdom of crowds lies in the fact that, although intrinsic divergence may exists among all participants, their collaborative opinion is still reliable. However, uncertainty remains pervasive across individuals due to the lack of authoritative opinions and adequate background information. Also, from a jurisdiction perspective, where most jurors are expected to bear nearly the same judgement in mind, uncertainty may rise from the objective difference of the accessibility to and processing of available information. Decision making for online users is such a case. According to various history and backgrounds of individual, we assume that one individual has a single error-rate $\epsilon_i$, where $\epsilon_i \in (0, 1)$, indicating the probability that this particular participant will make a conflicting judgement to the latent true value.

DEFINITION 4 (INDIVIDUAL ERROR RATE - $\epsilon_i$). *The individual error rate $\epsilon_i$ is the probability that a juror conducts a wrong voting. Specifically*

$$\epsilon_i = Pr(vote\ otherwise | a\ task\ with\ ground\ truth\ A)$$

Ground truth $A$ can be $0(false)$ and $1(true)$, which is unknown by the jury.

While utilizing the wisdom of crowds, another issue has to be scrutinized further: how reliable these distributed judgements are? In a voting, a subset of Jury may vote wrongly due to the reasons we listed above, but there are an exponential number of cases where different sets of jurors can make mistakes. So here we define the concept of *Carelessness* to capture such cases.

DEFINITION 5 (CARELESSNESS - $C$). *The Carelessness C is defined as the number of mistaken jurors in a jury $J_n$ during a voting, where $0 \leq C \leq n$.*

Informally, we define a possible combination of mistaken jurors in a Jury as *Minority*, and straightforwardly we can find that there are an exponential number of such possible *Minorities* for a particular $C$. To exactly measure the reliability of a crowd, we define the Jury Error Rate as the probability that a Voting $\vec{v}$ misses the true value because more than half of the jurors are wrong:

DEFINITION 6 (JURY ERROR RATE - $JER(J_n)$). *The jury error rate is the probability that the Carelessness C is greater than $\frac{n+1}{2}$ for a jury $J_n$, namely*

$$JER(J_n) = \sum_{k=\frac{n+1}{2}}^{n} \sum_{A \in F_k} \prod_{i \in A} \epsilon_i \prod_{j \in A^c} (1 - \epsilon_j)$$

$$= \Pr(C \geq \frac{n+1}{2} | J_n)$$

where $F_k$ is all the subsets of $S$ with size $k$ and $\epsilon_i$ is the individual error rate of juror $j_i$.

As shown in the definition, a naive method to calculate $JER$ is to enumerate all *Minorities* and accumulate their probabilities. This method entails an exponential number of product terms and renders any possible algorithm inefficient. We present two efficient algorithms to accelerate the computation in Section 3.1

Note that as a decision making problem, we assume that for each discerning task, there exists an objective and true judgement which is latent for all the participants before the crowd's decision is aggregated.

## 2.2 Crowdsourcing Models

In most crowdsourcing and human computing applications, how to motivate users to participate is an interesting problem. In this section, we present two models to describe the most prevailing phenomenon on crowdsourcing services.

### 2.2.1 Altruism Jurors Model

As cited in [12], people who spend a huge bunch of time online are not a uniform sample from the real world; on the contrary, white, educated people with middle and higher income are the main part of the online community. Among them, there exist plenty of altruistic workers who are motivated to participate in a task simply because they are interested or they feel they are obligated to participate. In this case, no matter how talented the worker is, he or she requires no extra payment as incentive, which means that any set of such users can form and function as a jury.

DEFINITION 7 (ALTRUISM JURORS MODEL - ALTRM). *While selecting a jury J from all candidate jurors (choosing a subset $J \subseteq S$), any possible jury is allowed.*

Table 3: Summary of Notations

| Notation | Description |
|---|---|
| $j_i$ | a juror (with index $i$) |
| $\epsilon_i$ | individual error rate of jury i |
| $J_n$ | a (candidate) jury with size $n$ |
| $C$ | the number of wrong jurors in a jury $J_n$ during a voting |
| $JER(J_n)$ | the probability that the jury $J_n$ fails under a voting |
| AltrM | Altruistic Jurors Model |
| PayM | Pay-as-you-go Jurors Model |
| $n$ | the size of a formed Jury $J_n$ |
| $N$ | the size of all candidate jurors |

Note that, by terming "allowed", we mean that this particular selection is legal and can be used to conduct a decision making task. Next, we are going to investigate the case that jurors ask for extra incentive, which may lead to a case that a jury is not "allowed".

### 2.2.2 Pay-as-you-go Model

The same thing happens on the U.S. court, that quite a portion of citizens do not feel glorious when selected to attend a trial or a hearing as a juror due to the loss of time and income, or simply because they feel intrinsically reluctant on particular issues. Meanwhile, most prevailing general purpose crowdsourcing platforms, like Amazon Turk, promise to pay the workers after they finish the task. With the monetary incentive, the jury selection procedure encounters more complicated problems. Financial incentive may also incur anchoring effect, which makes the aggregation of distributed knowledge even more sophisticated. However, at this stage, we only focus on the effect when a juror, or a set of them, are too expensive to be selected into a jury. Formally, we define the following model:

DEFINITION 8 (PAY-AS-YOU-GO MODEL - PAYM).
*While selecting a jury $J$ from all candidate jurors (choosing a subset $J \subseteq S$), each candidate juror $j_i$ is associated with a payment requirement $r_i$ where $r_i \geq 0$, the possible jury $J$ is allowed when the total payment of $J$ is no more than a given budget $B$, namely $\sum_{\forall j_i \in J} r_i \leq B$.*

## 2.3 Problem Definition

Here we formally define the Jury Selection Problem as an optimization problem.

DEFINITION 9 (JURY SELECTION PROBLEM - JSP).
*Given a candidate juror set $S$ with size $|S| = N$, a budget $B \geq 0$, a crowdsourcing model(AltrM or PayM), the Jury Selection Problem(JSP) is to select a jury $J_n \subseteq S$ with size $1 \leq n \leq N$, that $J_n$ is allowed according to crowdsourcing model and $JER(J_n)$ is minimized.*

For brevity, please refer to Table 3 as a collection of all the notations used in this paper.

## 3. JURY SELECTION ALGORITHM

Due to the existence of the variety of jurors' individual error-rates, it is non-trivial to form a best jury in terms of $JER$. An intuitive thinking might be that the best jury is selected from the best jurors, which means we can sort all the individuals with respect to their error-rates before selection. But how should we decide the size of a jury? As presented in the example of Section 1, the 5-juror group performs better than the 3-juror one, but when the size increases, a 7-juror group does not show any superiority over the smaller jury. Moreover, as we discussed in Section 2.1.2, even with a given set of candidate jurors, the calculation of JER is not trivial, let alone the optimal selection problem.

In this section, we will formally investigate calculation of JER, and then discuss JSP with *AltrM* model, along with two efficient algorithms. Then, we present a solution for JSP under *PayM* model and discuss its complexity.

### 3.1 Calculation of JER

To select the best jury with the minimum JER, we first have to calculate JER for a given jury. Theoretically, the number of jurors who give wrong votes on a task(the $C$ in Definition 5) is a random variable which follows the Poisson-Binomial distribution [21]. A naive method(used in the motivation example) to calculate this value is to enumerate all the *Minorities* and calculate the overall error-rate for each of them. Obviously this method is very inefficient and even impractical when the number of candidate jurors becomes large. Fortunately, we can speed up this calculation with dynamic programming.

#### 3.1.1 A Dynamic Programming Method

To simplify the illustration of calculating JER, we here assign an ordering $\{j_1, j_2, \cdots, j_n\}$ for the $n$ jurors(not necessarily sorted), and refer $J_m$ to the set of $\{j_1, j_2, \cdots, j_m\}$.

The basic observation is that there are repeated calculations of JER from a smaller jury to a larger one. Given a jury $J_n$ with size n, if $j_n$ makes a wrong vote(actually it can represent an arbitrary juror), the target $JER(J_n)$ becomes the probability that $\frac{n+1}{2} - 1$ jurors vote incorrectly in the jury $J_n \setminus \{j_n\} = J_{n-1}$. Straightforward enough, when this juror makes a correct decision, $JER(J_n)$ becomes the probability that still at least $\frac{n+1}{2}$ jurors are wrong, but in the smaller jury $J_{n-1}$ excluding $j_n$. Formally we have the following lemma:

LEMMA 1. *The calculation of JER of Jury with size n can be split into smaller ones:*

$$\Pr(C \geq L | J_n)$$
$$= \Pr(C \geq L-1 | J_{n-1}) \cdot \epsilon_n + \Pr(C \geq L | J_{n-1}) \cdot (1 - \epsilon_n)$$

*where*

$$\Pr(C \geq 0 | J_m) = 1 \quad \forall \quad 0 \leq m \leq n$$
$$\Pr(C \geq m | J_n) = 0 \quad \forall \quad m > n$$

PROOF. Straightforward from Definition 6 □

The initial conditions have a clear meaning: $\Pr(C \geq 0 | J_m) = 1$ covers all situations given a jury, and $\Pr(C \geq m | J_n) = 0$ means the number of wrong jurors cannot exceed the size of a given jury. Then, based on Lemma 1, we can propose the following method to calculate $JER(J_n)$.

We present a bottom-up implementation in Algorithm 1 by maintaining a two-dimension array $E[i, j]$ in Line 2: Starting from $\Pr(C \geq 1 | J_1) = \Pr(C \geq 0 | J_0) \cdot \epsilon_1 + \Pr(C \geq 1 | J_0) \cdot (1 - \epsilon_1)$, we can iteratively compute $JER$ with an increasing size of jury. Specifically, $\Pr(C \geq 1 | J_m)$ can be calculated from $\epsilon_m$ and $\Pr(C \geq 1 | J_{m-1})$ because all $\Pr(C \geq 1 | J_{m-1})$ is 0 by Lemma 1. After calculating from $\Pr(C \geq$ $1|J_1$) to $\Pr(C \geq 1|J_{n-\frac{n+1}{2}})$, we can further calculate all $\Pr(C \geq 2|J_m)$ from $\Pr(C \geq 1|J_{m-1})$ and $\Pr(C \geq 2|J_{m-1})$ in the same manner. In this way, we can finally obtain the value of $JER(J_n)$ after $\frac{n+1}{2}$ rounds.

**Algorithm 1** DP-based Algorithm
**Input:** A jury $J_n$
**Output:** The Jury Error Rate $JER(J_n)$
1: $m \leftarrow \lfloor (n+2)/2 \rfloor$;
2: create an array $E[0,\ldots,m][0,\ldots,n]$ with all value as 0;
3: **for** $i = 0 \leftarrow n$ **do**
4:   **for** $j = 1 \leftarrow \frac{n-1}{2} + i$ **do**
5:     **if** $i == 0$ **then**
6:       $E[i][j] \leftarrow 1$;
7:     **else**
8:       $E[i][j] \leftarrow E[i-1][j-1] * \epsilon_j + E[i][j-1] * (1-\epsilon_j)$;
9:     **end if**
10:   **end for**
11: **end for**
12: **return** $JER(J_n) \leftarrow E[n][\frac{n+1}{2}]$;

Note that in each iteration(a fixed size of $l$), we only need to calculate $(n - \frac{n+1}{2} + l = \frac{n+1}{2})$ times of JER because $\Pr(C \geq l|J_m)$ with larger $l$ is not necessary. Then we have following analysis.

COROLLARY 1. *The calculation of $JER(J_n)$ entails at most $O(n^2)$ time and at most $O(n)$ space using Dynamic Programming.*

PROOF. There are in total $\frac{n+1}{2}$ rounds of iteration, and within each iteration, there are at most $(n - \frac{n+1}{2} = \frac{n-1}{2})$ times of simple calculation as to Lemma 1. Each simple calculation entails $O(1)$ time cost, and thus the calculation of $JER(J_n)$ needs $O(\frac{n+1}{2} \cdot \frac{n-1}{2}) = O(n^2)$ time.

At any point, to calculate a particular $\Pr(C \geq l|J_m)$, only two vectors of previous calculated value are needed, i.e. one vector for the $JER$ with same $l$ and one vector for the ones with $l-1$. Hence, the space cost is $O(2 \cdot n) = O(n)$. □

### 3.1.2 A Divide and Conquer Method

Since the time complexity of the dynamic programming-based algorithm is $O(n^2)$, we have to spend much time in calculation when the size of jury is quite large, therefore we need to improve the efficiency of calculating JER further.

In this subsection, we propose another more efficient algorithm CBA(<u>C</u>onvolution-<u>b</u>ased <u>A</u>lgorithm), which is based on the divide & conquer framework instead of the dynamic programming strategy. In order to compute JER, it is equivalent to obtain the probability distribution of $C$, which is the number of jurors who give wrong votes on a task. The main idea of this algorithm is stated as follows: the algorithm first considers the probability distribution of $C$ as coefficients of a polynomial. Then, it divides the jury into two parts and recursively calls this process. When the jury has only one juror, the probability distribution of one juror is considered as the coefficients of a one-order polynomial. After partition, it uses the polynomial multiplication to merge the probability distributions of juries with smaller sizes and finally obtains the complete probability distribution of $C$. The process of divide & conquer will spend $O(n^2)$ time. However, we can use Fast Fourier Transform (FFT) method to speed up the process of polynomial multiplication. Thus, the final time complexity of CBA algorithm is $O(n \log n)$. The pseudo-code of CBA is shown in Algorithm 2.

In Algorithm 2, we first address the special case that there is only one candidate in $J_n$ in Lines 2-4. Then from Line 6 to 8, the algorithm divides the computation of $D_C$ into two parts, and in Line 9, the convolution-based merging is conducted via FFT. Note that the returned value in Line 11 is $D_C$, the distribution of C, in order to support recursive calling. $JER(J_n)$ can be easily retrieved as $\sum_{i=\frac{n+1}{2}}^{n} D_C[i]$.

**Algorithm 2** Convolution-based Algorithm(CBA)
**Input:** A jury $J_n$
**Output:** The $JER(J_n)$
1: **if** $n = 1$ **then**
2:   $D_C[0] = 1 - \epsilon_1$ ;
3:   $D_C[1] = \epsilon_1$ ;
4:   **return** $D_C$;
5: **else**
6:   Dividing $J_n$ into two parts: $J_{n1}$ and $J_{n2}$, where $|J_{n1}| = \lfloor \frac{n}{2} \rfloor$ and $|J_{n2}| = \lceil \frac{n}{2} \rceil$;
7:   $D_{C1} = CBA(J_{n1})$;
8:   $D_{C2} = CBA(J_{n2})$;
9:   $D_C =$ convolution of $D_{C1}$ and $D_{C2}$ via FFT;
10: **end if**
11: **return** $D_C$;

### 3.1.3 Lower Bound-based Pruning

Both the dynamic programming-based and convolution-based algorithms want compute JER efficiently. However, computing JER for each $J_n$ is redundant because there is only one jury which is finally selected. Thus, it is important to filter out insignificant candidate juries as early as possible. A natural idea is to quickly find a tight lower bound of JER to determine whether a new JER needs be computed. Based on the Paley-Zygmund inequality [26], we can obtain a tight lower bound of JER as follows.

LEMMA 2 (LOWER BOUND-BASED PRUNING). *Given a jury with size $n$, the lower bound of $JER(J_n)$ is shown as follows,*

$$JER(J_n) \geq \frac{(1-\gamma)^2 \mu^2}{(1-\gamma)^2 \mu^2 + \sigma^2}$$

*where $\mu = \sum_{i=1}^{n} \epsilon_i$, $\sigma^2 = \sum_{i=1}^{n}(1-\epsilon_i)\epsilon_i$, and $\gamma = (\frac{n+1}{2}/\mu) \in (0,1)$.*

PROOF. According to the definition of $JER$, $JER$ is the following probability:$\Pr\{C > \frac{n+1}{2}\}$, where $C$ is the number of jurors who give wrong votes on a task. Since $C$ is a random variable following Poisson Binomial distribution, the expectation and variance of $C$ are $\mu = \sum_{i=1}^{n} \epsilon_i$ and $\sigma^2 = \sum_{i=1}^{n}(1-\epsilon_i)\epsilon_i$ respectively.

Based on the Paley-Zygmund inequality, we can know: for a positive random variable $C$,

$$\Pr\{C \geq \gamma E(C)\} \geq \frac{(1-\gamma)^2 \mu^2}{(1-\gamma)^2 \mu^2 + \sigma^2}$$

where $E(C)$ means the expectation of the random variable $C$. Hence, let $\gamma = \frac{n+1}{2}/\mu$, we can rewrite the formula of JER as follows:

$$JER(J_n) = \Pr(C \geq \frac{n+1}{2}) = \Pr\{C \geq \gamma \cdot \mu\} \geq \frac{(1-\gamma)^2 \mu^2}{(1-\gamma)^2 \mu^2 + \sigma^2}$$

□

According to Lemma 2, we can observe that the time complexity of computing the lower bound of JER is only $O(n)$, where $n$ is the size of the jury. Thus, the time cost of lower bound calculation is smaller than that of both algorithms. Therefore, the lower bound-based pruning should improve efficiency of the algorithms computing JER.

## 3.2 JSP on AltrM

### 3.2.1 Monotonicity with Given Jury Size

Before we reach the final algorithm for selecting an optimal jury on *AltrM*, we firstly investigate whether *JER* follows monotonicity on individual error-rate with a given jury size.

LEMMA 3. *The lowest JER originates from the Jurors with lowest individual error-rate among the candidate jurors set $S$.*

PROOF. W.l.o.g, we pick one $j_i$ of the n jurors in a given Jury $J_n$ with size n. Then $JER(J_n)$ can be transformed as below:

$$JER(J_n) = \Pr(C \geq \frac{n+1}{2}|J_n)$$
$$=\epsilon_i(\Pr(C \geq \frac{n+1}{2} - 1|J_{n-1}))$$
$$+ (1-\epsilon_i) \cdot (\Pr(C \geq \frac{n+1}{2}|J_{n-1}))$$
$$=\epsilon_i(\Pr(C \geq \frac{n+1}{2} - 1|J_{n-1}) - \Pr(C \geq \frac{n+1}{2}|J_{n-1}))$$
$$+ \Pr(C \geq \frac{n+1}{2}|J_{n-1})$$
$$=\epsilon_i(\Pr(C = \frac{n+1}{2} - 1|J_{n-1}) + (\Pr(C \geq \frac{n+1}{2}|J_{n-1}))$$
$$=\epsilon_i \cdot A + B$$

It is obvious that $A = \Pr(C = \frac{n+1}{2} - 1|J_{n-1}) \geq 0$, so that the $JER$ is a monotone increasing function with respect to individual error rate $\epsilon_i$. In this way, given a jury with size $n$ and the candidate jurors set $S$ with size $N$, we finally prove Lemma 3 by contradiction:

Suppose a Jury $J'_n$ has the lowest $JER$, and juror $j'_i$ is one of the members but with a rank higher than $n$ in the candidate juror set $S$ in an ascending order with respect to the individual error rate $\epsilon$. Because $J'_n$ consists of $n$ jurors, there must be a juror $j_i$ which is not in $J'_n$ whose individual error rate $\epsilon_i$ is lower than that of $j'_i$. By substituting $j_i$ with $j'_i$ into $J'_n$, according to the monotone increasing property above, $J'_n$ will have a lower JER, which contradicts with its previous assertion as optimal. □

### 3.2.2 Algorithm for AltrM

Based on Lemma 1 and Lemma 3, we can now propose an efficient algorithm to solve *JSP* on *AltrM* model: firstly, the algorithm sorts all jurors in the candidate juror set $S$ in an ascending order of $\epsilon$; then varying the possible size $n$ of jury from 1 to $N$, we calculate $JER_{\mathbf{V}}(J_n)$; finally we return the jury with the lowest $JER$ as solution.

In the Line 5, the algorithm first checks the condition whether $\gamma$ is less than 1 according to Lemma 2. If $\gamma$ satisfies the condition, the algorithm then runs a lower bounding test in Line 6 as early-termination condition for JER. If $\gamma$ is larger than 1, the algorithm will calculate JER directly.

---

**Algorithm 3** Framework of *JSP* on *AltrM* - (AltrALG)

**Input:** A subset of candidate juror set $s \subseteq S$
**Output:** A subset of candidate juror set $S$ with lowest $JER$( Definition 6);
1: $s := j_1, j_c = j_1;$  //$j_c$ is the largest juror in current set
2: sort $\vec{\epsilon}$ in ascending order into $\vec{\epsilon}_{sorted}$;
3: **for** $n = 1 : N$ with step of 2 **do**
4:  form candidate Jury $J_n$ by selecting the first $n$ jurors in $\vec{\epsilon}_{sorted}$;
5:  **if** $\gamma(s \cup \{j_c \text{to} j_n\}) < 1$ **then**
6:   **if** $JER_{lowerbound}(s \cup \{j_c \text{to} j_n\}) \leq JER(s)$ **then**
7:    calculate $JER(s \cup \{j_c \text{to} j_n\})$;
8:    update $s$ accordingly;
9:   **end if**
10: **else**
11:   calculate $JER(s \cup \{j_c \text{to} j_n\})$;
12:   update $s$ accordingly;
13: **end if**
14: **end for**
15: **return** $s \subseteq S$ as proposed jury

---

Note that we assume that Algorithm 2 is called to calculate JER.

The time cost on Line 11 is $O(N \cdot \log N)$ according to Lemma 1, and there are in total $N$ times of iterations. The sorting in Line 2 costs $O(N \cdot \log N)$ time, and comparison in Line 16 costs $O(1)$ time. Hence the algorithm for *JSP* on *AltrM* model has time complexity of $O(N \cdot \log N \cdot N) = O(N^2 \cdot \log N)$.

The algorithm for JSP on *AltrM* is denoted as AltrALG for simplicity.

## 3.3 JSP on PayM

In *PayM* model, each candidate juror is associated with a requirement $f_i$, and the Jury Selection Problem is about how to select the best jurors within a limited budget. In such a setting, a candidate Jury may be rejected because of excessive requirement of payments. And we will discuss how to estimate the expected payment of each candidate juror in Section 4.2.

### 3.3.1 NP-hardness

Compared to a traditional 0/1 Knapsack Problem(KP), JSP on *PayM* features JER as an objective function, instead of a simple summation of values of the selected items. Although we have proved in Lemma 3 that the JER is lowest when selecting individuals with the lowest error-rates, given a fixed size of jury, we do not know its monotonicity with respect to the size of a selected jury. These properties make the objective function a generally non-linear one, which shows much more hardness than the general Knapsack Problem. The general 0/1 Knapsack Problem is a classic NP-complete problem [14], and we reduce one of its variant, the $n$th-order Knapsack Problem(nOKP), to the JSP problem.

LEMMA 4. *JSP on* PayM *is NP-complete*.

PROOF (SKETCH OF PROOF OF LEMMA 4). We reach the proof of Lemma 4 by proving the np-completeness of its decision version, the Decision JSP(DJSP), i.e. given a JSP instance and a value $v$, decide whether a Jury $J_n$ can be

1500

selected so that $JER(J_n)$ is equal to $v$. According to Definition 6, which is the objective function of JSP, this optimization problem is a $n$th-order Knapsack Problem. We then follow the proof of NP-hardness of Quadratic Knapsack Problem(QKP)given by H. Kellerer, et al. in [14] to prove the hardness of nOKP.

A $n$th-order Knapsack Problem(nOKP) is a Knapsack problem whose objective function has the form as follow:

$$optimize \quad \underbrace{\sum_{i_1 \in n} \sum_{i_2 \in n} \ldots \sum_{i_n \in n}}_{n} V[i_1, i_2, \ldots, i_n] \cdot x_1 x_2 \ldots x_n$$

where $V[i_1, i_2, \ldots, i_n]$ is a $n$-dimensional vector indicating the profit achieved if item $[i_1, i_n, \ldots, i_n]$ are selected simultaneously.

Given an instance of traditional KP, we can construct an $nOKP$ instance by defining the profit $n$-dimensional vector as $V[i, i, \ldots, i] = p_i$ and $V[otherwise] = 0$ for all $i$, where $p_i$ is the profit in traditional KP. The weight vector and objective value remain the same. □

### 3.3.2 Approximate Algorithm

Because of the complexity of JSP on $PayM$, we present a heuristic algorithm to tackle this problem with best efforts. The underlying idea of Greedy Heuristic is to sort all the candidate jurors according to the product of their error rate and requirement, i.e. $\epsilon_i \cdot r_i$. Then we increase the size of a jury from 1 to $N$ with a growing pace of 2. Each time when the enlargement still comply with the budget constraint, we allow this enlargement after validation of improvement on JER. The difference between this algorithm and the traditional greedy algorithm for 0/1 Knapsack Problem is that when the algorithm considers a new candidate, not only the weight, but also the benefit should be compared. This is also the reason why JSP on $JER$ is harder than traditional KP. Formally we present the Greedy Heuristic Algorithm in Algorithm 4:

---

**Algorithm 4** Framework of $JSP$ on $PayM$ - (PayALG)
**Input:** A set of $N$ candidate jurors $S$ with the vector of individual error-rates $\vec{\epsilon}$ and the vector of requirements $\vec{r}$, and a non-negative budget $B$
**Output:** A subset of candidate juror set $s \subseteq S$
1: sort $\epsilon_i \cdot r_i$ in ascending order into $\vec{j} = \{j_1, j_2, \ldots, j_N\}$;
2: $\overline{r} := 0$, $s := \emptyset$, $pair := 0$;
3: **while** $r_i > B$ **do**
4:     increase $i$ in $\vec{j}$;    //find the first $j_i$ in $\vec{j}$, s.t. $r_i \leq B$;
5: **end while**
6: select $j_i$, $s := s \cup \{j_i\}$ ;
7: update accumulated requirement$\overline{r} := r_i$;
8: **for** $m = i + 1 : N$ **do**
9:     **if** $pair = 0$ and $r_m + \overline{r} \leq B$ **then**
10:      set $j_m$ as pair, $j_{pair} := j_m, r_{pair} = r_m$;
11:      set pair flag $pair := 1$;
12:     **else if** $pair = 1$ and $r_m + r_{pair} + \overline{r} \leq B$ and $JER(s \cup \{j_{pair}, j_m\}) \leq JER(s)$ **then**
13:      select $j_m$ and its pair, $s := s \cup \{j_{pair}, j_m\}$;
14:      set pair flag $pair := 0$;
15:     **end if**
16: **end for**
17: **return** $s \subseteq S$ as the proposed jury

---

Note that in Line 1, $\overline{r}$ is the current accumulated requirement. Due to the requirement of odd size of a jury, the greedy algorithm considers a pair of candidate jurors as one enlargement. Then in Line 6, we find the first feasible juror whose requirement is less budget $B$. And in each step, a *pair* flag will be set(in Line 10) to indicate that one more candidate should be admitted to examine the updated JER.

The algorithm for JSP on $PayM$ is denoted as PayALG for simplicity.

## 4. PARAMETER ESTIMATION

In this section we will further discuss several possible approaches to estimate the individual error-rate $\epsilon_i$ and expected payment requirement $r_i$ in $PayM$ model from micro-blog service data.

Note that to obtain person's individual error-rate and expected cost is itself an emerging research topic, in the tide of extending power of crowdsourcing from AMT to more general platforms. Our work focuses on forming up best crowd and aggregating answers, which is a fundamental step of this trend. For complete illustration of the proposed frame work, we propose a method to infer the requirement from the age of an account. In fact, any other reasonable measures can be smoothly plugged in to our framework.

### 4.1 Estimate Individual Error-rate

In this subsection, we propose a possible method to estimate error-rates according to their authority in terms of knowledge where decision is made. Basically, our approach is to construct a user-graph for Twitter data according to their forwarding operation retweet "RT", and ranking users in the constructed graph. Each user is then assigned a ranking score, or confidence score, which represents the quality of the user and can be directly translated to an error-rate. Details are explained as follows.

#### 4.1.1 Graph Construction

The Twitter social network is modeled as a graph $G(V, E)$, where $V$ is the set of nodes, each of which represents a user, and $E$ is the set of edges. Instead of making use of the "following-and-follower" user relationship on Twitter, we link two nodes or users based on their *retweet* actions. A *retweet* action is that a user quotes or re-broadcasts another user's tweet. Intuitively, the more a user's tweets are retweeted by other users, the more authoritative or influential the user is. Previous work [5] has adopted retweet measurement for influence analysis on Twitter. It is also indicated that mainstream news organizations and celebrities are the major groups of people who often induce a high level of retweet actions. Therefore, by building a retweet-relationship-based graph and ranking users in the graph, we can identify reliability or quality of users to a large degree.

More specifically, we define an ordered-pair of users ($user_1$, $user_2$) if $user_1$ has ever retweeted $user_2$'s tweets, which we call a retweet-relationship pair. In our Twitter data, a tweet containing "RT @$username$" indicates that $username$'s tweet is retweeted, where $username$ is any legal username on Twitter. There are two possible cases that suggest the existence of a retweet-relationship pair in our Twitter data:

1. A tweet $t$ released by a user $user_1$ contains one and only one substring "RT @$username$"

2. A tweet $t$ released by a user $user_1$ contains more than one substrings "RT @$username$"

where $username$ is any legal username.

In the first case, let $user_2$ be the user with $username$ in the substring "RT @$username$", and then $(user_1, user_2)$ is a retweet-relationship pair. In the second case, let $user_2$, $user_3$, ... $user_N$ be the users whose usernames are contained in the substrings "RT @$username$" and that appear in the order appearing in the tweet $t$. As a prototype, this $t$ indicates a retweet-relationship chain: $user_N$ is the original author, $user_{N-1}$ retweetes $user_N$'s tweet, ..., and $user_1$ retweets $user_2$'s tweets, which is released as $t$. For this retweet-relationship chain, we extract N-1 retweet-relationship pairs $(user_1, user_2)$, $(user_2, user_3)$, ..., and $(user_{N-1}, user_N)$.

In the set of retweet-relationship pairs we find in our Twitter data, we link $user_1$ to $user_2$ once and only once for each pair $(user_1, user_2)$, which results in a directed user-graph.

---

**Algorithm 5** Graph Construction

**Input:** Tweets dataset $T$. Each record $r(t, author)$ includes the tweet's content $c$ and author $author$
**Output:** A directed graph $G(V, E)$
1: Set $V = \emptyset$
2: Set $E = \emptyset$
3: **for** each $r(t, author) \in T$ **do**
4:   $last\_user = author$
5:   Add $last\_user$ to $V$
6:   **while** $c$ contains substring $str$='RT @[\w]+[\W]+' **do**
7:     Extract username $user\_retweeted$='[\w]+' from $str$
8:     Add $user\_retweeted$ to $V$
9:     Add $edge(last\_user \rightarrow user\_retweeted)$ to $E$
10:     Delete $str$ from $t$
11:     $last\_user = user\_retweeted$
12:   **end while**
13: **end for**
14: **return** $G(V, E)$;

---

### 4.1.2 User Ranking

In order to measure quality of users, we need to rank users in the graph constructed. Popular webpage ranking algorithms HITS [15] and PageRank [22] have been applied to solve expert location problems in social networks [29]. Since our constructed graph is also a directed and connected user network that is suitable to run graph-base ranking algorithms, we also employ HITS and PageRank on the user-graph to obtain quality or confidence scores of users.

We can obtain authority scores and hub scores for users by employing HITS. We adopt the authority scores as quality scores. The page rank scores calculated by PageRank are directly used as quality scores.

We generalize the framework of estimating user scores by HITS in Algorithm 6 and by Pagerank in Algorithm 7. We find in the real dataset that most top ranking users discovered by Pagerank overlaps with the ones identified by HITS.

### 4.1.3 Error-rate

Due to the Power law distribution characteristics of social network users, and also for the ease of differentiating the qualities among all candidate users, we normalize the score

---

**Algorithm 6** Quality Score Calculation with HITS

**Input:** A directed graph $G(V, E)$
**Output:** Quality scores $Score$ for each $user \in V$
1: Initialize $Score$ and $Hub$ to 1
2: **while** iteration not ends **do**
3:   Reset $Score$ to 0
4:   **for** each $edge(u, v) \in E$ **do**
5:     $Score[v] = Score[v] + Hub[u]$
6:   **end for**
7:   Normalize $Score$
8:   Reset $Hub$ to 0
9:   **for** each $edge(u, v) \in E$ **do**
10:     $Hub[u] = Hub[u] + Score[v]$
11:   **end for**
12:   Normalize $Hub$
13: **end while**
14: **return** $Score$;

---

**Algorithm 7** Quality Score Calculation with PageRank

**Input:** A directed graph $G(V, E)$
**Output:** Quality scores $Score$ for each $user \in V$
1: Set damping factor $d$
2: $n = |V|$
3: **for** each $user \in V$ **do**
4:   $Score[user] = \frac{1}{n}$
5:   $Out[user] = |\{v|edge(user, v) \in E\}|$
6:   $In\_Set[user] = \{u|edge(u, user) \in E\}$
7: **end for**
8: **while** iteration not ends **do**
9:   **for** each $user \in V$ **do**
10:     $New\_Score[user] = \frac{1-d}{n} + d \sum_{i \in In\_Set[user]} \frac{Score[i]}{Out[i]}$
11:   **end for**
12:   Copy $New\_Score$ to $Score$
13: **end while**
14: **return** $Score$;

---

of each user to range in $(0, 1)$ as follows, where $\alpha$ and $\beta$ are normalization factors(setting are given in Section 5.2):

$$\epsilon_i = \beta^{-\alpha(score_i - min)/(max - min)}$$

where $min$ and $max$ are the minimum and maximum score values obtained from Algorithm 6 and Algorithm 7.

## 4.2 Integrated Cost Estimate in *PayM* Model

There are several works related to user profiling and community inferring on social networks [30], from all kinds of information like online behaviors [20] or even user names [28]. Based on these attributes of users, we can imply the taste and preference of users [17].

The task of further determining the individual requirement for each user is a domain-specified procedure, and needs careful designs according to different types of tasks to be proposed. The detail of such mechanism is out of the scope of this work, and we propose an optional indicator to estimate the individual requirement $r_i$.

**Inferring from Account Age**

Here we propose to use a single attribute as the indicator of individual requirement: the age of a user account since registration. We roughly assume that, the more experienced a user is, the less he or she will be interested in a task.

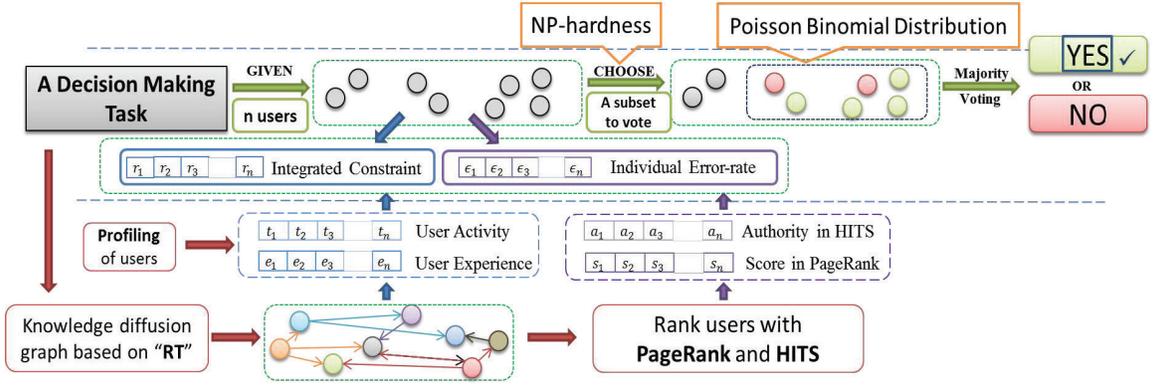

Figure 2: System Overview

In Section 4.1.2, after selecting a candidate set $S$ according to the individual error rate, we retrieve the age $t_i$ of each user from his or her registration date. The value of individual requirement can be estimated as follow:

$$r_i = \frac{(t_i - min)}{max - min}$$

where $min$ and $max$ are the extremum values of the estimated account age of all users.

## 4.3 System Overview

Although this study mainly focus on the modeling and algorithmic solution of the practical jury selection problem, we briefly present a conceptual system overview for better illustration in Figure 2.

There are mainly two parts in the system, one is for estimating individual error rate and requirement for a large set of candidates, and the other is concerning about selecting the best crowd. As illustrated in the upper part of Figure 2, a subset of candidates are selected to form a jury, and this jury will achieve a final Yes/No decision via Majority Voting scheme. For different situations, different parameter estimation methods should be utilized to best capture the candidates' characteristics.

## 5. EVALUATION

In this section, we present our experimental evaluation of the performance of AltrALG and PayALG, as well as an experimental study of the JSP problem, namely the relationship among individual error rates, the optimal jury size and given budgets.

In Section 5.1, we utilize synthetic datasets to evaluate the performance of both algorithms, which follow the normal distributions with varying mean values and variance values. In Section 5.2, we retrieve candidate juror data from real micro-blog service data(Twitter) by following the algorithms described in Section 4.

All the experiments are conducted on an Intel(R) Core(TM) i7 3.40GHz PC with 8GB memory, running on Microsoft Windows 7.

## 5.1 Synthetic Data

To simulate individual error rates and requirements without bias, in this section we produce synthetic datasets following normal distributions with varying mean values and variance values. JSP characteristics are investigated both on $AltrM$ and $PayM$ models. Then we evaluate the efficiency of AltrALG on $AltrM$ model and the effectiveness of PayALG on $PayM$ model.

### 5.1.1 Evaluation on AltrM

**JSP Traits on $AltrM$**

$AltrM$ model can actually be interpreted as one special case in $PayM$ where all the requirements are zero or a unlimited budget $B$ is given. In such case, the only concern of JSP is to determine the size of a jury whose JER is minimized.

The synthetic dataset is generated as following: we generate 1,000 candidate jurors, whose individual error rates follow a normal distribution with mean values varying from 0.1 to 0.9, and variance values from 0.1 to 0.3. We then perform AltrALG on this dataset and record the performance as shown in Figure 3(a). In this figure, $var$ means variance of the individual error rates.

It is straightforward to interpret our findings: when most of the candidates are reliable, namely whose individual error rates are less than 0.5, the optimization problem is conducted as searching in a very flat slope. This causes a randomized distribution of a best jury size as shown in the left shoulder of curves in Figure 3(a). On the other hand, when most of the individuals are error-prone, which means individual error rate is larger than 0.5, a best jury has to reduce its size to keep the jury as "the hands of the few"[1]. In addition, under this synthetic dataset, the threshold of reducing the jury size is around the point where the mean of individual error rates is 0.5. This actually implies the turning point where wisdom of crowd may malfunction.

**Efficiency on $AltrM$**

Because the AltrALG can always find the optimal solution, we then mainly evaluate its efficiency with a growing data size. Specifically, we track the running time of AltrALG with an increasing input size. In this setting, we generate dataset of individual error-rates with mean value of 0.1, and vary the size of candidate jurors from 2,000 to 6,000 with variance of 0.05 and 0.1 respectively. The results are shown in Figure 3(b). The line denoted by m(0.1) means the dataset is with variance of 0.1 and the algorithm is conducted without lower-bounding checking in Line 6 of AltrALG; the line with legend of m(0.1, b) means the lower-bounding checking is conducted normally as in AltrALG.

---
[1] The case where "truth rests in the hands of a few."

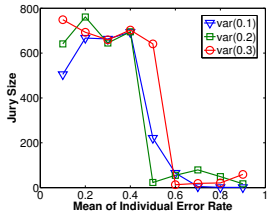
(a) Jury Size v.s. Individual Error-rate

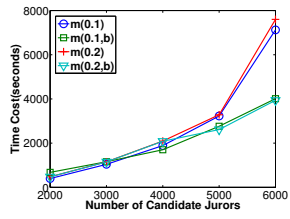
(b) Efficiency of JSP on AltrM

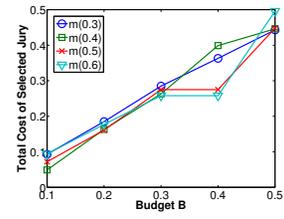
(c) Budget v.s. Total Cost

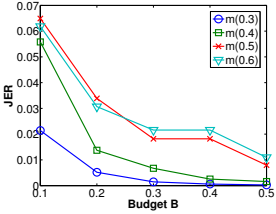
(d) Budget v.s. JER

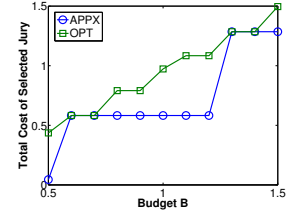
(e) APPX v.s. OPT on Total Cost

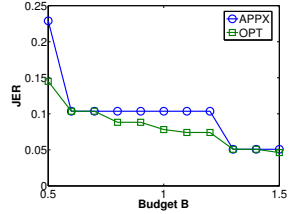
(f) APPX v.s. OPT on JER

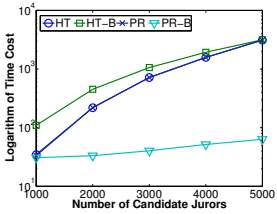
(g) Efficiency of JSP on Twitter Data

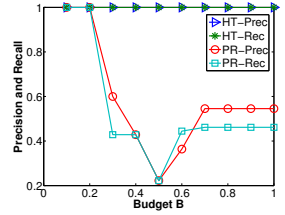
(h) Precision & Recall on Twitter Data

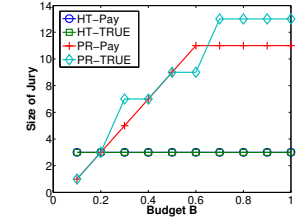
(i) Jury Size on Twitter Data

Figure 3: Experiments Results

It can be noticed in Figure 3(b) that when the data size is small(2000 to 3000), the enhancement of algorithm by checking lower bounding entails unfortunately more running time than the non-enhanced one, which is mainly caused by the overhead of checking the condition of lower-bounding pruning. When the size grows, we may easily generalize that the running time of the enhanced algorithm increases slower than the non-enhanced one with a ratio of $O(1/\log N)$.

### 5.1.2 Evaluation on PayM

Since the greedy heuristic algorithm on *PayM* runs with a linear time cost, instead of focusing on the efficiency issue, in this subsection we investigate the relationship between the quality of a selected jury and the given budget. Moreover, we evaluate the quality of our selection algorithm by comparing JER and the total cost with ground truth.

**JSP Traits on *PayM***

JSP on *PayM* is a classic situation where most crowdsourcing applications are conducted, and the influence of budgeting is one of the essential factors in this setting. Thus, we investigate the relationship among the budget posed and the resulted JER, as well as the final cost.

We generate a candidate jurors set with individual error rate mean of 0.2, variance of 0.05 and set size of 1,000;the individual requirement is generated from the normal distribution with mean value of 0.4, 0.5 and 0.6 respectively , variance value of 0.2. The given budget $B$ varies from 0.1 to 0.5 and the results are shown in Figure 3(c) and Figure 3(d). The line with $m(0.3)$ as legend represents the performance of jurors with mean error-rate of 0.3.

The results in Figure 3(c) again verifies the findings in Section 5.1.1 that for jurors with individual error rate of more than 0.5, the algorithm tends to reduce the size of the selected jury but pay higher for each selected juror. From Figure 3(d), we can generalize that a raising budget can improve jury quality by reducing JER, and a candidate set with lower individual error-rates(e.g. the one of m(0.3)) forms a better jury within same budget.

**Effectiveness on *PayM***

By terming "Effectiveness", we are to investigate the discrepancy between the results obtained by PayALG and the ground truth. Due to the NP-hardness of JSP on *PayM*, we calculate the ground truth via enumerating all possible combinations of jurors and check whether a combination achieves the lowest JER while satisfying the budget requirement. Since the running time increases exponentially with a growing size of candidates in this enumeration method, we generate a candidate jurors set with size of only 22. The error-rates of these candidates follows the normal distributions with mean of 0.2 and variance of 0.05 and 0.1 respectively; the individual requirement is also following a normal distribution with mean of 0.05 and variance of 0.2. We vary the budget from 1 to 3 with step of 0.2, and the results are shown in Figure 3(e) and Figure 3(f). In the legend, "APPX" represents the results from Algorithm 4, and "OPT" represents the ones from ground truth.

The results from ground truth in Figure 3(e) show that budget is indeed the constraint of forming better jury. In Figure 3(f), its shows that the heuristic PayALG achieves the optimal JER as ground truth 4 times out of 11. Moreover, the biggest discrepancy appears with the lowest budget $B = 0.5$, and with an increasing budget, the JER given by PayALG is getting closer to the one of ground truth, because a larger budget loosens the constraint of forming a better though sub-optimal jury.

## 5.2 Real Twitter Data

The dataset we use in this section is a previously published collection of public twitter time-line messages, recording random samples gathered in two days. We estimate individual error rate $\epsilon_i$ and $r_i$ from the data based on the methods in Section 4, which use HITS and PageRank to estimate the error-rate of each users. The normalization of individual error-rates follows equation in Section 4.1.3. There are in total 689,050 nodes but since most of them have very sparse mutual 'RT' relationship, so we simply choose the 5,000 users with highest scores.

### 5.2.1 Evaluation on AltrM

For online services, efficiency is an important issue to be considered. To evaluate whether the proposed algorithm is a practical technique, we test it on both datasets generated from HITS and PageRank. Both the datasets are normalized according to the equation in Section 4.1.3 with parameter $\alpha = 10, \beta = 10$. We evaluate the running time by varying size of candidate jurors from 1,000 to 5,000. The results are shown in Figure 3(g). In the legend, "HT" stands for data from HITS and "PR" stands for data from PageRank, and "-B" stands for the results achieved with the lower-bounding enhancement in Line 6, AltrALG.

As shown in Figure 3(g), the running time of the algorithm without bounding enhancement on two datasets is almost the same. But with bounding enhancement, the running time of the algorithm on data set generated by PageRank(PageRank data in short) is largely reduced while that of HITS increases. This is due to the difference between the two datasets, that after normalization, a larger portion of users in RageRank data has error-rates close to extremes(0 or 1) than the ones in HITS do. This distribution makes more users in the PageRank data satisfy the condition of using bounding enhancement: $\gamma \in (0, 1)$ according to Lemma 2 and thus avoid the unnecessary calculation of JER. But for users in HITS, the overhead of checking condition for lower bounding entails even more time cost.

### 5.2.2 Evaluation on PayM

We evaluate the performance of the approximation algorithm on both HITS and PageRank datasets. Due to the power law distribution of online user's error-rates, the size of the best jury converges quickly and the values of JER is thus reduced to 0. Thus in this subsection, we focus to providing a precision and recall value of the approximation algorithm. As previously mentioned, the ground truth comes from enumeration of all possible combinations and thus entails exponentially increasing time cost. Thus we retrieve top 20 candidates via both HITS and PageRank algorithm, and their error rates are normalized according to the equation in Section 4.1.3 with $\alpha = 10$ and $\beta = 10$. To provide meaningful budget testing variables, we vary the budget B as $0.1\%M, 1\%M, 10\%M$ and $20\%M$, where $M$ is the average value of estimated requirement of all candidate users multiplied by the number of candidate jurors. We present the precision and recall values in Figure 3(h). In the legend, "-Prec" stands for precision values and "-Rec" stands for recall values.

It can be seen in Figure 3(h) that results from HITS data have precision and recall with 1, but the results from PageRank have lower resemblance with ground truth in terms of precision and recall. As we discussed in Section 5.2.1, there are a relatively larger number of jurors in PageRank who have low error-rates than the ones in HITS, and this broadens the feasible solution space for forming a jury and in turn brings forward the low precision and recall values. In addition, as shown in Figure 3(i), the size of jury formed on PageRank data is close to the one from ground truth; and the size of jury formed on HITS always identical to ground truth. However, despite such low precision and recall value, the JER given by PayALG is still low enough(0.00075) for as a credible jury.

## 6. RELATED WORK

**Crowdsourcing** Research on crowdsourcing overlaps with several other topics like social computing, human computing, collective/collaborative intelligence, etc. It provides a new problem-solving paradigm[2, 18] and has branched into several areas. In database community, new types of queries are developed to aggregate distributed knowledge. [19] proposes "Qurk" to manage crowdsourced tasks as in relational database. [10] propose "CrowdDB" to organize human intelligence to solve problems that are naturally hard for computers.

[23]considers the situation where humans are invited to enhance a graph search procedure, and proposes an algorithm to find the optimal target nodes for crowd participation. Thus the work in [23] has certain resemblance with the problem studied in the sense of improving crowdsourcing performance. Other works have also provided creative usages of wisdom of crowd in multimedia annotation [6] and document searching [1].

**Worker Quality** As illustrated in the previous sections, crowdsourcing applications succeed when enough problem-solvers are well organized and their efforts are harvested intelligently. However, the overall quality of all individual workers is also important for the quality of final output. A well endorsed work by I., Panagiotis [13] proposes to use soft labels to improve the quality of a task finished by crowds, for that soft label can differentiate spam workers and bias workers. The work in [25] provides a Bayesian model to use maximum likelihood for inferring error rates of crowdsourcing workers.

In the application of data sourcing from the crowd, [7] proposes to use Markov Chain Monte Carlo to estimate the error rate of the data from crowdsourcing activities. Hence, work related to data cleaning can also be considered to reconcile the confliction within a crowd [11].

In this paper, we discuss the relationship between individual worker quality and the quality of one special product, the reliability of voting. And one major difference is that on AMT task requesters cannot choose worker actively; however, to ask question on micro-blog service is intrinsically equipped with the "@" markup, which enables the selection of workers.

**Expert Team Formation** Another application resembles the problem of finding a suitable set of users is Expert Team Formation Problem [16]. In Expert Team Formation Problem, the task normally has a specific requirement for certain skills which are possessed by different candidate experts. At the same time, cost of choosing one expert is also defined, e.g. communication cost or influence on personal relationship etc. Therefore the Team Formation problem is to minimize the cost while fulfilling the skills requirement.

Besides using explicit graph constraints, Expert Team Formation problem based on communication activities is also studied. In [3, 8], emails communication is used for expertise identification.

## 7. CONCLUSION

In this paper, we study the Jury Selection Problem(JSP) for decision making tasks on micro-blog services, whose challenges are calculating JER and finding the optimal subset under a limited budget. We explicitly discuss the formation of such probability and propose two efficient algorithms to calculate it within $O(n^2)$ and $O(n \cdot \log n)$ time respectively.

Models of altruistic users($AltrM$) and of incentive-requiring users($PayM$) are proposed to capture characteristics of crowdsourcing applications. The $AltrM$ model features the monotonicity of JER on individual error rate, and JSP on $AltrM$ model is NP-hard. We propose an efficient algorithm for JSP on both models.

We verified the proposed algorithms on both synthetic and real datasets through extensive experiments.

## 8. ACKNOWLEDGEMENTS

This work is supported in part by the Hong Kong RGC GRF Project No.611411, National Grand Fundamental Research 973 Program of China under Grant 2012-CB316200, HP IRP Project 2011, Microsoft Research Asia Grant, MRA11EG05 and HKUST RPC Grant RPC10EG13.